\documentstyle[pra,aps,preprint]{revtex}
\usepackage{rotate,epsf}
\begin{document}
\author{Fedor S. Kilin}
\address{
Belarus State University, Skaryna Ave. 4, 220050 Minsk, Belarus 
Fedia@dragon.bas-net.by} 
\title{   Analysis of Investment Policy in Belarus }
\date{\today}
\maketitle
\begin{abstract} 
The optimal planning trajectrory is analysed on the basis of the growth model  
with effectiveness. The saving per capital value has to be rather high initially  
with smooth decrement in the future years. 
 \end{abstract}

\section{Introduction}
         For the development of the long-term economic strategy and for
realization of appropriate investment policy one needs the theoretical model
of economic growth.  All developed countries pay a considerable attention to
researching of such theoretical models, as well as to developing of
corresponding instrumental means that are important for calculation of the
concrete prognoses and programs~[1] .The practical experience, accumulated by
many countries, demonstrates that the most effective tool for the development
of strategic directions of an economic policy on a long-run period are the
special economic-mathematical models of small dimension (macromodels). Such
models are elaborated on the base of the theory of economic growth~[2].

 In traditional macromodels the principal attention is payed to the estimation
of future dynamics of the investments that determine trajectories of the
economic growth. However, economic growth depends not only on scales of
investment resources, enclosed in economics: these trajectories are also
determined by a number of the so-called quality factors~[3]. Moreover, the
final economic growth results depend basically on these factors.

The orientation of economy to the extensive growth by escalating of wide area
investments only cannot ensure an achievement of the useful final results
without paying attention to the quality factors. The history of development of
many countries, including the former USSR, shows the possibility of economical
development by the principle of production for the sake of
production. Economical system absorbs huge investment resources and augments
volumetric parameters.  Nevertheless this type of investment policy is not
able to raise essentially standards of living.

The limited prospects of the economic development on the base of only
extensive factors are demonstrated in the Solow model~[4]. The estimation of
influence of technical progress is made on the basis of the Solow model with
technical progress~[2]. This model accounts for the contribution of technical
progress in the simplest way. It is based on the rather relative concept of
autonomous progress advance. The intensity of the autonomous progress effect
on production growth is completely determined by the time factor. A
quantitative estimations of such effect are received on the basis of the
production functions method. The parameters for this method are calculated by
means of econometric processing of dynamic rows that define change of
production volumes. A row of successive numbering of time periods (years,
quarters or other periods) is used as a dynamic row, that corresponds to the
change of technical progress.

In essence another approach to define the
quality factors contribution to the growth of production volumes has been
realized in growth model with effectiveness.  This approach does not use
insecure parameters of rather relative econometric models.  It is based on a
direct estimation of the change of an economic system material capabilities
that are indispensable for the realization of social purposes.  The growth
model with effectiveness is most adequate for simulation of economic growth in
the present economical situation for the Republic of Belarus. The main purpose
of this work is to formulate the optimal planning problem for the growth model
with effectiveness and solution of this problem: such solution is especially
important for researches of economic growth problems, because it answers the
main question that appears when choosing the investment policy. The essence of
this question consists in choosing between maintenance of current demand
(consumption) and maintenance of the future demand (capital investment). The
solution of the optimal planning problem allows to point such investment
policy, at which the economic system works at best. The formulation of the
depends on the purposes that the economic system has.  In this work one takes
the most realistic version of such purpose - maximization of the welfare
integral which is consumption per man during the modeled period of time.

The article is organized as follows: section II deals with the concept of
production effectiveness. Growth model with effectiveness is described in
section III.  In section IV one considers optimal planning problem
formulation. Maximum principle is applied in section V. Solution of the optmal
planning problem is analyzed in section VI. Policy options are considered in
section VII.  Main achievements of this work are briefly concluded in section
VIII.
    
\section{Concept of production effectiveness} 
The growth model with effectiveness uses modern approach to predict the
economic growth trajectory and define the quality factors contribution in
increasing of production volumes. This approach does not rest on
accident-sensitive parameters of rather conditional econometric models, but it
is grounded on the direct estimation of those changes of material capabilities
of an economical system that are necessary for realization of the chosen
purposes. On the basis of such estimation it is expedient to draw a conclusion
about the advance, reached in economics: one admits the existence of "advance"
only in the case, when the capabilities for implementation of the social and
economic purposes are extended. At the same time, from the standpoint of the
target approach, opening capabilities scales shows the degree of the obtained
advances.

Within the framework of such an approach there is a more successful name for
the identification of the advance, reached in economics. This name is
"increase of production effectiveness" (or, accordingly, its reduction, if the
capabilities for the implementation of the social and economic purposes are
reduced). This term is used in this work to reflect all quality factors
influence on economic growth . Thus, the primary task of this chapter is to
design a criteria index that should characterize change of production
effectiveness on the macrolevel. Such an index should be used as one of the
most important variables of the dynamic macromodel, which reflects the
relationship of this macromodel with other macroeconomic indexes that describe
the intensity of production and use of resources, as far as accumulation and
non-productive consumption. Within the framework of the dynamical macromodel
of economical growth, the submodel of the production effectiveness has been
developed from the standpoint of the target approach~[5].  The economical
system under consideration is supposed to be closed in the sense that the
total amount of the needs is fulfiled at the expense of the production
only. The system does not move its production into other countries in the debt
or gratuitously (that does not eliminate barter with an environment on the
equivalent basis for the coordination of commodity pattern of production with
consumption pattern).

To define an effectiveness criterion of production it is necessary to reveal
the certain economical form of the operational outcome of the economical
system and resources, used for its achievement. For this purpose one needs to
define not only spatial - organizational, but also temporal boundaries of this
system in order to reveal and to agree final output indexes with the initial
input factors involved in the process of production.

If we abstract from the natural and external economical factors and consider
the effectiveness of an economical system, without taking into account its
temporal boundaries, then the labour force is the only used resource, and the
amount of created material benefits and services intended only for the
non-productive consumption is the outcome of the system operation.  In this
case it would be possible to define the effectiveness index on the basis of
the simple comparison of the total final consumption and used labour force.
It is always necessary to associate an estimation of effectiveness with the
particular time frame, therefore time also should play a role of the factor
that limits the frameworks of the economical system. The inputs and outputs of
the system reflect not the connection with the environment.  The evolution of
the system depends on the past and defines the future evolution of the
system. The relations of the resources reproduction costs to their volumes are
the relevant characteristics of the reproduction process. For example the
consumption per man $\psi$ reflects the level of workers material
benefits. Any change of this index substantially determines the dynamics of
all indexes of the population welfare.  The relation of gross investment $S$
to a volume of accumulated productive capital $K$
\begin{equation}
\omega  = \frac{S}{K} 
\end{equation}
 predetermines the rate of capital growth in a decisive measure. The
 estimations $\psi$ and $\omega$ are connected with the indexes of labour
 productivity $p$ and capital per man $r$ as follows
\begin{equation}
p = \psi  + \omega  \cdot r. 
\end{equation}
This ratio is a consequence of the balance identity $Y = C + S$, where
$Y$ - output, $C$ - consumption, which originates from the assumption about
closure of the considered economical system (see fig.1).

To derive the equation (2) from the balance identity, one has to divide it by
the labour force volume $L$ and take into consideration that
\begin{equation}
{S \over L} 
           = 
                {\textstyle{S \over K}} \cdot {\textstyle{K \over L}}  
           =
                 \omega  \cdot r. 
\end{equation}
Relative resource estimations 
$\psi$  
 and  
$\omega$   
 can be accepted as objective functions  that  describe the implementation  
degree of two main  purposes  of  the  reproductive  process. 
These purposes form one general purpose of an economical system. Here we
denote it as "integral purpose".  It implies the maximization of satisfaction
of current and future needs of society.  The index of effectiveness should
serve as the objective function that permits to estimate quantitatively the
integral purpose implementation.

For the concrete definition of the integral
purpose and the integral index of effectiveness it is necessary to distribute
the priorities between two introduced primary purposes. This distribution
leads to their coordination and resolving of the contradiction between the
primary purposes. Such priorities are reflected quantitatively at distribution
of the gross internal product to parts intended for reproduction of two kinds
of manufacturing resources. The purposes of the society define the proportions
of such distribution and concretize the effectiveness index formula that
should serve as the integral object function of the reproduction process.

Below all variables are counted for the given year. Here we assume that during
t-th year the usage of the manpower quantity
$L_t$ 
 and capital  of  the volume 
$ K_t$ 
 makes the gross internal product of the volume 
$Y_t$. 
The corresponding  values of the relative indexes  
of labour productivity and   capital  per man   are peer 
$p_t$ 
 and  
$r_t$, respectively. 
The indexes  
$p_t$   
 and  
$r_t$  
 are the most  important  quantitative characteristics of  the process of  
reproduction as generalized   technology on   the  macrolevel. 
Though their values  
don't determine base-line values of  the object functions 
$\psi_t$ 
 and 
$\omega_t$ 
 uniquely, 
but  they limit the area of their possible  values. In accordance with  
the general limiting condition the  above  indicated  main  balance identity reads 
\begin{equation} 
p_t  = \psi _t  + \omega _t  \cdot r_t.  
\end{equation}

The particular values of estimations 
$\psi_t$ 
 and  
$\omega_t$ 
 can be derived from the area, 
restricted  by  the equality (4).
This area depends   
on the distribution of the made gross internal product, which   arrests the norm of  
accumulation  
$\delta _t$. 
The last is defined as the fraction of the gross  accumulation in the total  
amount of the gross internal product 
\begin{equation} 
\delta _t  = \frac{{S_t }}{{Y_t }}. 
\end{equation}
At the given norm of accumulation and the certain labour productivity level as
well as capital per man, the values of object functions are determined
uniquely according to the formulas
\begin{eqnarray}
\psi _t  = (1 - \delta _t ) \cdot p_t, \nonumber\\  
\omega _t  = \frac{{\delta _t  \cdot p_t }}{{r_t }}. 
\end{eqnarray}
Their values  are the starting point for the analysis of the effectiveness  
dynamics.

Here we consider the values of variables for the next year $t+1$. 
The capital per  
man  increases to the value of 
$r_{t + 1}$. 
The labour productivity for the given  capital  per  man  
reaches the  value  
$p_{t + 1}$. 
The dynamics of the reproduction process effectiveness characterizes the
change of the implementation capabilities of two main purposes that have to be
reflected in the change of area of acceptable values of the object functions
$\omega$ 
 and 
$\psi$. 
In order to estimate the  degree of the indicated change quantitatively one  
needs to compare labour productivity in current year with  the  value 
$\psi _t  + \omega _t  \cdot r_{t + 1}$,  
which describes  that minimal  production volume per  man. This allows to  
keep the values  of the object functions  
$\omega$  

 and 
$\psi$  
 at the level  of the basic year. 
 \begin{equation}
p_{t + 1}  = \psi _t  + \omega _t  \cdot r_{t + 1}.  
\end{equation}
        The last corresponds to the equal effectiveness of production in
current and basic years. It means that the values of both object functions can
be saved at a basic level and both can not be increased at once. Note that
actual values of estimations 
$\omega $ 
 and 
$\psi$  
 in one year $t+1$ can  differ from  
$\psi _t$  
 and 
$\omega _t$, 
but  one of them will be   more basic, and another  will  be less.

Assume the labour productivity level in one year 
$t + 1$ 
 surpasses the bound 
\[
p_{t + 1}  > \psi _t  + \omega _t  \cdot r_{t + 1}.
\] Therefore in this year there appears the possibility for simultaneous
increase of values of two object functions in contrast to the year $t$.  As
mentioned before these object functions describe the level of implementation
of the main purposes of the reproduction process. It gives the ground to
suppose that in one year $t + 1$ the value of the integral object function is
augmented and, therefore, the level of efficiency is raised. Moreover, the
value of a difference $p_{t + 1} - \psi _t - \omega _t \cdot r_{t + 1}$ allows
to judge that there appeared capability to increase the values of object
functions $\omega$.  Therefore this difference can serve as the characteristic
of the effectiveness increase degree.

The actual increment of labour
productivity $\Delta p_t = p_{t + 1} - p_t$ can be decomposed in two parts
\begin{eqnarray}
\Delta _1  &=& \psi _t  + \omega _t  \cdot r_{t + 1}  - p_t,   \\
\Delta _2  &=& p_{t + 1}  - \psi _t  - \omega _t  \cdot r_{t + 1}.
\end{eqnarray}
The value $\Delta _1$ represents the increment of the labour productivity, at
 which the level of effectiveness remains invariable. It can also be
 interpreted as the increment of theproductivity reached at the expense of the
 extensive increase of the capital per man (at the basic level of the
 effectiveness ). Another part of the increment of labour productivity $\Delta
 _2$ quantitatively characterizes the additional capabilities of
 implementation of the main purposes of production appeared in one year $t +
 1$.
We    start   from  the reason that the growth of  the effectiveness  is  the  only   
source of increase of such capabilities. Therefore it is possible to consider the value $\Delta _2$
  as the increment of the productivity  reached at the expense of the increase of  
efficiency. 
The  ratio of  the  increment 
$\Delta _2$ 
 to the base-line value of the labour productivity 
\begin{equation}
\frac
{{\Delta _2 }}
{{p_t }} 
               = 
                   \frac
                       {{p_{t+ 1}  - \psi _t  - \omega _t  \cdot  r_{t + 1} }}
                       {{p_t }}. 
\end{equation}
represents  the  rate of the productivity increment at the expense of the effectiveness  
increase. This ratio  can be identified with  the  rate  of   the  increment of  the  
effectiveness   of production.

In equation (10) we replace $p_{t + 1}$ with $p_t + \Delta p_t$ and $r_{t + 1}
$ with $r_t + \Delta r_t $ and simplify it using the balance identity (2).
The rate of the increment of effectiveness depends on the pure increments of
the labour productivity $\Delta p_t $ and capital per man $\Delta r_t $
\begin{equation}
\frac
  {{\Delta _2 }}
  {{p_t }} 
                  = 
                       \frac
                          {{p_{t + 1}  - \psi _t  
                          - \omega _t  \cdot  r_{t + 1} }}
                          {{p_t }}.
\end{equation}
The equation (11) can be rewritten using  the  absolute  increases of the  
corresponding indexes to the  rates of their change. 
For this purpose one substitutes  
the intensity of reproduction of the capital from the formula (6) into equation (11): 
\begin{eqnarray}
i_\varphi   
         &=& 
               \frac{1}{{p_t }} 
               \cdot 
               \left(
                 \Delta p_t 
               - \frac
                      {{\delta _t  \cdot p_t}}
                      {{r_t }} 
                 \cdot 
                 \Delta r_t 
               \right) \nonumber \\
        &=& 
               \frac{{\Delta p_t }}
                    {{p_t }} 
            - \delta _t  
              \cdot  
              \frac{{\Delta r_t }}
                   {{r_t }}. \nonumber 
\end{eqnarray}                   
Here we introduce the notations for rates of increment of the corresponding
indexes $i_P = \frac{{\Delta p_t }}{{p_t }}$ and $i_r = \frac{{\Delta r_t
}}{{r_t }}$.  Then the estimation of dynamics of a production effectiveness
takes the form
\begin{equation} 
i_\varphi   = i_p  - \delta  \cdot i_r  
\end{equation}

Thus, the conducted analysis results in a conclusion that the reference point
for the society that seeks to achieve the social and economic purposes, can't
serve the increase of labour productivity, which traditionally was considered
to be the main criteria index in our economics. The obtained two equivalent
dynamical formulas of the index of a production effectiveness (11) and (12)
demonstrate that for the estimation of outcomes of the economical development
it is necessary to correct the growth of labour productivity allowing for the
change of the capital per man. The bigger part of the made gross internal
product is routed to the reproduction of the productive capital, the more is
the deviation of the dynamics of the criteria index of effectiveness from
dynamics of labour productivity. The last can serve as the main reference
point of control only in that case, when the capital per man remains
invariable. In this work we consider the usage of the effectiveness index as
the preferred reference point of control.

\section{Growth model with effectiveness} 
The following six main macroeconomic indexes have been selected as variables
of the Growth model with effectiveness $r$ - capital per man, $p$ - output per
man, $\psi$ - consumption per man, $\omega$ - saving per capital, $\lambda$ -
sum of amortization rate and population rate, $i_\varphi$ - the rate of
economics effectiveness.

For simplification of model and its analysis the
following preconditions are adopted: the number of workers in economics
changes with constant rate of increment $n$, (so that the dynamics of
employment can be recorded with the help of the function $L = L_0 \cdot
e^{nt}$).  The amortization rate $\beta$ is also invariable.  Besides the
differential form of notation is used. This form is more convenient for
solution and analysis and widely used in the literature on
economic-mathematical modeling~[5]. For conversion to the differential form
one supposes, that the values of the considered economical indexes are
continuously differentiable functions of time. The increments of indexes per
unit of time, selected as a step for the analysis, are substituted by
derivative from functions that describe their dynamics.

 The main equation of the model is based on the definition of the
effectiveness index. The formula (11), derived earlier for the estimation of
dynamics of effectiveness and introduced in the incremental form, can be
converted to the following equivalent equation in the differential form
\begin{equation}
\dot p(t) 
          = 
             i_\varphi  (t) \cdot p(t) 
           + \omega (t) \cdot \dot r(t). 
\end{equation}
The differential equation (13) is included in the developed macroeconomic
model and plays the role of production function, which links indexes of
capital per man and labour productivity. Besides, the model includes the main
balance identity (2)
\begin{equation} 
p(t) = \psi (t) + \omega (t) \cdot r(t). 
\end{equation}                        
The ratio (6) can be rewritten in differential form as follows 
\begin{equation} 
\omega (t) 
\cdot r(t) 
          = 
             \delta (t) 
             \cdot p(t). 
\end{equation} 
The differential equation for 
$r$ comes  from the equation of capital dinamics  
\begin{equation}
dK = S(t) - \beta  \cdot K(t).
\end{equation}
To pass from the increment of capital to the index of  the increment of   capital   
per  man,  it is  necessary  to  differentiate  the formula  for  capital  per man 
\begin{eqnarray}
dr 
  &=&
       d\left( 
             {\frac{{K(t)}}{{L(t)}}} 
        \right) \nonumber \\
  &=& 
       \frac
            {{dK \cdot L(t) - K(t) \cdot  dL}}
            {{L(t)^2 }} \nonumber \\
  &=&  \frac
            {{dK}}
            {{L(t)}} 
     - \frac
            {{K(t)}}
            {{L(t)}} 
     \cdot  
       \frac
            {{dL}}
            {{L(t)}}.  
\end{eqnarray} 
Here we introduce the population rate  
$n = \frac{{dL(t)}}{{L(t)}}$.
The last assumption allows to rewrite equation (17) in the following form:  
\begin{eqnarray} 
dr 
   &=&
        \frac
             {{S(t) - \beta  \cdot K(t)}}
             {{L(t)}} 
       - n \cdot r(t) \nonumber \\ 
   &=&  
       \frac
            {{S(t)}}
            {{L(t)}} 
      - (\beta  + n) 
        \cdot r(t). 
\end{eqnarray}
Now in (18) instead of 
$\frac{{S(t)}}{{L(t)}}$ 
  we  put equivalent expression from (3), and for the sum of two  
constants 
$n + \beta$ 
 we enter new identification  
$\lambda  = n + \beta$. 
Then it is possible to enter into the model  the  equation, that reflects  the   
correlation between  capital  per  man and saving per capital.  
\begin{equation} 
\dot r(t) = (\omega (t) - \lambda ) \cdot r(t). 
\end{equation}
Equations (13) - (15)  and  (19) form the set of four equations, which describes  
the  correlations  between six  abovelisted main macroeconomic indexes.

\section{Optimal planning problem formulation} 
It is nessesary to emphasize that this model is represented by six variables
and four equations. Therefore, the obtained set of equations (13)-(15),(19) is
incomplete.  In order to close the set of equations one introduces control
variables that describe investment policy. It is possible to do by assuming
$i_\varphi = const$ and $\omega (t)$ is a control variable. Then one can
formulate optimal planning problem. First, it is necessary to construct a
target functional. The task of the central planning establishment is to select
a feasible trajectory $\omega (t)$ that is the optimum for achievement of some
economic target.  The economic target of central planning organ should be
based on the standards of living, estimated by the consumption level. In
particular, it is presumed that the central planning organ has an utility
function, which determines utility $U$ at any moment of time as a function of
consumption per man $U = U[\psi (t)]$.  We assume that the utility function is
doubly differentiable and that the marginal utility is positive and
non-increasing function. Therefore, the utility function is concave and
monotonically increasing function. The utility function determines utility in
a certain year. However, the problem of the central planning organ is to
select the whole trajectory of consumption per man. For this purpose it is
necessary to compare indexes of utility that correspond to different
years. Suppose, that utilities in different years do not depend on each
other. The utility in any year does not depend on either consumption or
utility in any other year directly.  Utilities for different years are
supposed to be additive. This assumption is based on the fact, that the
proximate consumption is more important, than distant. We suppose that the
norm of discounting $d$ is constant and positive. The bigger norm of
discounting testifies the greater preference of the utilities closest in
time. Suspecting the exponential behavior of the discounting, we receive the
value of utility in year $t$ equal to $ e^{ - d(t - t_0 )} U[\psi (t)]$.
During the indicated time period from $t_0$ to $t_1$ the welfare $W$,
corresponding to the pathway of consumption per man $\psi (t)$, is determined
by integrating of all instantaneous utilities over the whole interval. The
time planning horizon $t_1$ can be finite or infinite.  In case this time is
finite, it is necessary to set the minimally acceptable value of capital per
man in final year to ensure the possibility of consumption outside the given
horizon of time. Here we try to avoid difficulties, connected with the
definition of the minimal value of capital per man in a final moment of
time. Consider, that $t_1$ is indefinite, so the control trajectory is
selected for all times in the future.  

However, in this case the welfare
integral can miss. The convergence of an integral is guaranteed, if the
following conditions are satisfied: the initial value of capital per man is
less than the maximal accessible level $\tilde k$ and the norm of discounting
is positive. In this case $c(t) \le f(\tilde k)$
 and
\begin{equation}
\int
\limits_{t_0 }^\infty  
      {e^{ - d(t - t_0 )} U(\psi (t))dt \le } 
\int
\limits_{t_0}^\infty  
      {e^{ - d(t - t_0 )} U(f(\tilde k))dt}  
      = 
       \frac
             {{U(f(\tilde k))}}
             {d}. 
\end{equation}
So the integral of welfare is bounded above.

We choose $r(t)$ and $p(t)$ as
state variables. We express consumption per man through control and state
variables from (14). Then we receive optimal planning problem for model with
effectiveness
\begin{eqnarray}
W 
  &=& 
       \int
           \limits_0^\infty  
            e^{ - d \cdot t} 
            U(p - \omega  \cdot r)dt 
      \to 
         \mathop{\max }
         \limits_{\omega (t)},  \nonumber  \\ 
\dot r 
   &=&
        (\omega  - \lambda )r,  \nonumber \\ 
\dot p 
   &=&
         pi_\varphi   
      + \omega 
       (\omega  - \lambda ) 
        \cdot r,  \nonumber \\ 
r(0) 
   &=& 
       r_0,   \nonumber \\ 
p(0) 
   &=&
       p_0.   
\end{eqnarray}

\section{Applying of  the Maximum principle} 
 
In this section we derive the equation for the saving per capital trajectory.  
When solving optimal planning problems with the help of  the Maximum principle,  
for each coordinate of state variables vector  
$x$  
 a costate variable is used. The  Hamiltonian function  is 
$H(x,y,u,t) = I(x,u,t) + yf(x,u,t)$, where 
$I(x,u,t)$ 
 stands for  the integrand of a target functional,  
$f(x,u,t)$ 
  for the vector of motion equations  right parts,  $u$  for control. 
We find functions $u(t)$,  
$x(t)$,   
$y(t)$ 
 that satisfy  the  following conditions 
\begin{eqnarray}
\mathop {\max }\limits_u H(x,u,y,t),  
t_0  
   &\le& 
       t, \\ 
\dot x 
   &=&
       \frac
            {{\partial H}}
            {{\partial y}},  
x(t_0 ) 
    =
        x_0, \\
\dot y 
   &=&
        - \frac{{\partial H}}{{\partial x}}, \\
H 
   &=&
         e^{ - d(t - t_0 )} 
         U(p - \omega  \cdot r) 
       + y_r (\omega  - \lambda )r 
       + y_p      
         \left[
          pi_\varphi   
        + \omega 
          (\omega  - \lambda )r
         \right].
\end{eqnarray}             
where 
$y_r$, 
$y_p$   
 are costate variables. 
Finally, by using of Eqs. (22) - (24) 
we can write the problem to be solved in the following form
\begin{eqnarray} 
0
   &=& 
        - e^{ - d \cdot t} U'(p - \omega r) + y_r  + 2y_p \omega  - y_p \lambda, \\ 
\dot 
   r 
   &=& 
        (\omega  - \lambda )r,  \\ 
\dot 
   p 
   &=& 
        pi_\varphi   + \omega (\omega  - \lambda )r,  \\ 
\dot y_r  
   &=& 
        \omega  
        \cdot e^{ - d \cdot t} 
        U'(p - \omega  \cdot r) 
      - y_r (\omega  - \lambda ) 
      - y_p 
        \omega (\omega  - \lambda ),  \\ 
\dot 
y_p  
    &=&  
      - e^{ - d \cdot t} 
        U'(p - \omega  \cdot r) 
      - y_p i_\varphi.
\end{eqnarray}
The system of five equations with five unknown functions is then obtained.
Four equations of this system are differential, one is algebraic. Solution of
this system of equations is equivalent to the solution of the problem (21).

The set of equations (26) - (30) has the solution for any kind of the utility
function. 
However in this work we perform the analysis of the elementary case
$U(z) = z$. 
In this case 
Eqs.~(29) and (30) for costate variables $y_r$ and $y_p$
do not depend on state ones $r$ and $p$.
In order to find an optimal savings trajectory $\omega(t)$
we have to solve only three equations (26), (29), and (30).
After that one can find state variables $r$ and $p$.
As a first step we obtain from Eq.~(26) an explicit formula for $\omega(t)$
written as a function of costate variables $y_r$ and $y_p$:
\begin{equation}
\omega  = \frac{\lambda }{2} + \frac{{\chi  - y_r }}{{2y_p }}, 
\end{equation}
where $\chi=exp(-dt)$ is an exponential function.
Note, that the solution of eq.(30) for $y_p$ can be written in the analytical form  
\begin{equation}
y_p (t) = y_p (0)e^{ - \mu  \cdot t}  + \frac{1}{{\mu  - d}}(e^{ - \mu  \cdot t}  - e^{ -  
d \cdot t} ), 
\end{equation}
where 
$\mu  = i_\varphi$.
Having   differentiated  the left and right parts of eq. (31),  one arrives at 
\begin{equation} 
\dot 
\omega  
       =  
         - \frac{1}{2}
           \frac{{(\dot y_r  - \dot \chi )y_p  
                - (y_r  - \chi )\dot y_p  
                 }}
                 {{y_p^2 }}.  
\end{equation}
By rewriting Eq.~(31) in the following form
\begin{equation}
\frac
  {{y_r  - \chi }}
  {{y_p }} 
             = 
                 \lambda  - 2\omega. 
\end{equation}
and substituting it in the right part of Eq.~(33), we find
\begin{equation}
\dot 
\omega  
      =  
        - \frac{1}{2}
          \frac
                {{(\dot y_r  - \dot \chi )}}
                {{y_p }} 
        + \frac{1}{2}
          \frac
               {{\dot y_p }}
               {{y_p }}
          (\lambda  - 2\omega ). 
\end{equation}
Then we     replace $\dot y_r$
 and 
$\dot y_p$ 
 in the right part of eq. (29) and (30) respectively
with the intermediate result
\begin{equation} 
\dot 
\omega  
         =  
           - \frac{1}{2}
             \frac{{[\omega \chi  - y_r (\omega  - \lambda ) - y_p  
                    \omega (\omega  - \lambda ) - \chi ]}}
                  {{y_p }} 
          + \frac{1}{2}
            \frac{{( - \chi  - y_p \mu  )}}
                 {{y_p }}
            (\lambda  - 2\omega ). 
\end{equation}  
Finally, we substitute the ratio of costate variables 
 \begin{equation}
\frac{{y_r }}{{y_p }} = \lambda  - 2\omega  + \frac{\chi }{{y_p }}. 
\end{equation}                               
found from Eq.~(34) in the equation (36) and obtain the following differential
equation for saving per capital 
\begin{equation} 
\dot 
\omega  
       =  
         - \frac{1}{2}\omega ^2  + (\mu  + \lambda  + \frac{\chi }{{y_p  
}})\omega  - \frac{{\lambda ^2 }}{2} - \frac{{(2\lambda  + d)\chi }}{{2y_p }} -  
\frac{{\mu \lambda }}{2}.
\end{equation}
Found equation together with Eq.~(32) is equivalent to the optimal planning
problem (21).
Solution of this equation gives an optimal savings trajectory $\omega(t)$.

\section{Optimal Savings Trajectory} 
 
The  differential linear equation (38) is  Riccati  equation which can be
solved by standard methods.. For the  
parameters values  
$\mu  = 0.015$,
$d = 0.2$ 
 (corresponding  to decreasing significance of consumption in e  times  for 5  
years), 
$\lambda  = 0.02$, 
and the initial value 
$\omega (0) = 0.05$, 
the solution of the Riccati equation (38) has been found numerically by the  
Runge-Kutta method. 
The obtained solution is  shown in  figure 2.

The equation (38) enables analytical consideration in   the  asymptotic
limit when
$\exp(-(d-\mu)t)\to 0$, i.e. when $d>\mu$
and 
$t >> 1/(d-\mu)$.
In  this  case  
we can neglect those terms in Eq.~(38) proportional to $\chi/y_p$
with a result
\begin{equation} 
\dot 
\omega  
         =  
             - \frac{3}{2}\omega ^2  
             + \mu \omega 
             - \frac
                    {{\lambda  (\lambda  + \mu )}}
                    {2}
         =
             - \frac12
               (\omega-\omega_-)
               (\omega-\omega_+),
\end{equation}
where $\omega_\pm=(\lambda+\mu)\pm\sqrt{\mu(\lambda+\mu)}$.
Note, that this equation is known as equation of interacting masses. 
        Having  integrated  it we obtain an asymptotic
optimal trajectory
\begin{equation}
\omega 
(t) 
         = 
            \frac
                 {
                   \omega_-
                   \exp{(-t\sqrt{\mu(\lambda+\mu)})}
                  +
                   c \omega_+
                 }
                 {
                  \exp{(-t\sqrt{\mu(\lambda+\mu)})}+c
                 }
\end{equation} 
where 
the constant $c$ is defined by the values $\omega_\pm$
and the value of $\omega(t_\infty)=\omega_\infty$
taken as ``initial'' value for the asymptotic regime of evolution
\[
c=(\omega_\infty - \omega_-)(\omega_+ - \omega_\infty).
\]
As it follows from obtained solution an asymptotic trajectory will be slightly
increased if $\omega_- < \omega_\infty < \omega_+$,
being restricted by the value of
 $\omega_+=(\lambda+\mu)+\sqrt{\mu(\lambda+\mu)}$.
An asymptotic behavior of state variable $r$
is obtained from Eq.~(27) by taking into account Eq.~(40)
\begin{equation} 
r 
   =
       \tilde r
       \exp{((\mu+\sqrt{\mu(\lambda+\mu)})t)}
       \left(
             \exp{((-\sqrt{\mu(\lambda+\mu)})t)}
             +c
       \right)^2,
\end{equation}
where 
$\tilde r$ 
  is a rate-fixing constant.

\section{Policy Options} 
 
There are two main applications of this analysis : recommendations to the
government for investment policy development and prediction of a long-run
period macroeconomic system development.

To define recommended investment volume in the current year, we need to obtain
the stock of capital statistical data for this year and to find the optimal
saving per capital value using the optimal planning problem solution. Then we
can evaluate the investment volume as a product of $K$ and $\omega$.  The next
year we change the optimal planning problem parameters to improve the accuracy
of calculation. The procedure of recommended investment volume evaluation
remains invariable. Thus this method allows the optimization of the investment
policy during any period of time.

Variables from economic growth models play important role in models of other
important fields of economic life too. A financial programming method combines
these models . The optimal planning problem for the model with effectiveness
can serve as a part of financial programming systems ~[7]. The solution of the
optimal planning problem may be useful for evaluating financial programming
system parameters.

To predict long-run period macroeconomic system development we put optimal
saving per capital trajectory into the growth model with effectiveness and
then calculate all macroeconomic variables from this model. Table 1 shows
predictions based on the Republic of Belarus economic parameters. Optimal
saving per capital trajectory for this model is the result of the optimal
planning problem solution from this work.

Table 1 shows that this variant of macroeconomic strategy allows the extension
of productivity potential by increasing the capital per man by 72.4~\% and the
output per man by 55.5~\%. The total growth of consumption per man is 51.6~\%.

\section{Conclusions} 
 
The main result of this article is the optimal investment trajectory for
Belarus for the period of 2000-2020. The number of additional results may be
of use for the investment policy development in other countries and other
periods. The nonlinear system of equations allows to analyze dependencies
between trajectories of main macroeconomic variables. The asymptotic solution
shows long-run perspectives of macroeconomic system. 
Equation from this
work may be used in financial programming systems.

A possible improvment of the present model is connected with the splitting of
capital
in two parts - private industrial capital
and public overhead capital.
Besides, a more realistic utility function can be chosen for the model.
The results of the inprovements will be published elsewhere.

\section*{Acknowledgement} 
Useful advises of professor Komkov are gratefully  
acknowledged.

\newpage
\begin{figure}[p]\centering
   \caption{Schematic representation of the production process }
\end{figure} 
\begin{figure}[p]\centering
   \caption{Numerical solution of the optimal planning differential equation. Parameters represent the  
economy of Belarus. Numerical solution shows that the  saving per capital ratio has to be  
rather high initially with smooth decrement in future years. This is the main result of this  
work}
\end{figure} 

\newpage
 
\begin{table}[]
  \begin{center}
    \caption{Republic of Belarus economic development variant, when the  
effectiveness rate is 1.5\% }
\begin{tabular}{|c|c|c|c|c|c|c|c|c|}
Year
&\multicolumn{2}{c|}{Capital per man}
&\multicolumn{2}{c|}{Consumption per man}
&\multicolumn{2}{c|}{Output per man}
&\multicolumn{2}{c|}{Savings per capital}\\
\cline{2-9}
&value&rate \%& value&rate \%& value&rate \%& value&rate \%\\
\hline
2000&780&  &111,0&   & 150&   & 0,050&   \\
2001   &
803,6 &
103,0 &
113,3 &
102,1 &
153,5 &
102,3 &
0,049 &
99,3 \\
2002 &
827,7 &
106,1 &
115,6 &
104,1 &
157,0 &
104,6 &
0,048 &
98,6 \\
2003 &
852,1 &
109,2 &
118,0 &
106,3 &
160,6 &
107,0 &
0,047 &
98,0 \\
2004 &
877,1 &
112,4 &
120,4 &
108,5 &
164,2 &
109,5 &
0,046 &
97,4 \\
2005 &
902,4 &
115,7 &
122,9 & 
110,7 &
167,9 &
112,0 &
0,046 &
96,8 \\
2006 &
928,3 &
119,0 &
125,5 &
113,0 &
171,7 &
114,5 &
0,045 &
96,2 \\
2007 &
954,6 &
122,4 &
128,1 &
115,4 &
175,6 &
117,1 &
0,045 &
95,6 \\
2008 &
981,4 &
125,8 &
130,8 &
117,8 &
179,5 &
119,7 &
0,044 &
95,1 \\
2009 &
1008,7 & 
129,3 &
133,5 &
120,3 &
183,6 &
122,4 &
0,044 &
94,6 \\
2010 &
1036,5 &  
132,9 &
136,3 &
122,8 &
187,6 &
125,1 &
0,043 &
94,1 \\
2011 &
1064,8 & 
136,5 &
139,2 &
125,4 &
191,8 &
127,9 &
0,043 &
93,7  \\ 
2012 &
1093,7 & 
140,2 &
142,1 &
128,0 &
196,1 &
130,7 &
0,042 &
93,2 \\
2013 &
1123,0& 
144,0 &
145,1 &
130,7 &
200,4 &
133,6 &
0,042 &
92,8 \\
2014 &
1153,0  & 
147,8 &
148,2 &
133,5 &
204,8 &
136,6 &
0,041 &
92,4 \\
2015 &
1183,5& 
151,7 &
151,3 &
136,3 &
209,4 &
139,6 &
0,041 &
92,0 \\
2016 &
1214,5& 
155,7 &
154,6 &
139,2 &
214,0 &
142,6 &
0,040 &
91,6 \\
2017 &
1246,1& 
159,8 &
157,9 &
142,2 &
218,6 &
145,8 &
0,040 &
91,2 \\
2018 &
1278,4& 
163,9 &
161,2 &
145,3 &
223,4 &
149,0 &
0,040 &
90,9 \\
2019 &
1311,2& 
168,1 &
164,7 &
148,4 &
228,3 &
152,2 &
0,039 &
90,5 \\
2020 &
1344,6& 
172,4 &
168,3 &
151,6 &
233,3 &
155,5 &
0,039 &
90,2 
     \end{tabular}
    \label{tab:2}
  \end{center}  
\end{table}

\end{document}